\documentclass[12pt,a4paper,twoside,alpha-refs]{wiley-article}

\usepackage{amsmath} 
\usepackage{siunitx} 
\usepackage{enumitem} 
\usepackage{hyperref}
\usepackage{cleveref} 
\usepackage{graphicx} 
\usepackage{amsfonts,amsmath,amssymb}
\usepackage{bm}
\usepackage[english]{babel}

\usepackage{comment}
\usepackage{float}
\hypersetup{
    colorlinks=false,
    linkcolor=blue,
    filecolor=magenta,      
    urlcolor=cyan,
}
\usepackage{geometry}
\papertype{Proceeding}
\paperfield{}

\title{\textbf{\huge{Atmospheric \& geodesic controls on muon rate: a numerical study \\based on Corsika}}}
\date
\abbrevs{}

\author[1]{A.~Cohu}
\author[2]{M.~Tramontini}
\author[3]{A.~Chevalier}
\author[1]{J.-C.~Ianigro}
\author[1*]{J.~Marteau}

\affil[1]{Institut de Physique des 2 Infinis de Lyon (IP2I), IN2P3, CNRS, Université Lyon 1, UMR 5822}

\affil[2]{CONICET - Facultad de Ciencias Astronómicas y Geofísicas, Universidad Nacional de La Plata, La Plata, Argentina}
\affil[3]{MUODIM, 31 rue Saint-Maximin, 69003 Lyon}

\corraddress{J. Marteau PhD-HDR, IP2I Lyon, IN2P3, CNRS, Université Lyon 1, UMR 5513, 69622 Villeurbanne, France}
\corremail{marteau@ipnl.in2p3.fr}

\runningauthor{\LARGE{\textit{A.~Cohu~et~al.}}}

\begin{document}
\fontsize{10}{13}\selectfont

\maketitle

\begin{abstract}
Muon rate models play a key role in converting measured data into direct informations from detector calibration to density images extraction. A given experiment requires a proper modelization of the muon flux taking into account its position and atmospherics conditions. Two approaches are commonly used~: either through semi-empirical models calibrated at best on existing data and extrapolated to the field experiment conditions or via Monte-Carlo simulations. The latter offer the advantage to tackle down in a single approach all parameters, such as barometric conditions, geomagnetic field, atmosphere density etc. All these approaches remain less reliable when dealing with muons close to the horizon, a major motivation that triggers the present study based on the CORSIKA simulation framework. 
\end{abstract}

\section{Introduction}

Applications of atmospheric cosmic rays (CR) have grown in numbers in the last decade in the field of muography. Measurements of the muons flux attenuation or deviation have been successfully applied to the imaging or monitoring of large geophysical, archaeological or industrial structures. Among all charged particles reaching ground level, muons are the most numerous. Their energy loss or scattering when crossing a given length of matter is exploited to reconstruct the density of the medium. The measurement of the muon energy is usually not possible in small and compact field detectors, which are standard trackers using particle physics techniques (scintillators, resistive plate chambers, micro-megas etc). 
The simulation of the muon flux follows in general two different approaches : 
\begin{itemize}
    \item \textit{Analytical models} which provide semi-empirical formula adjusted on experimental datasets. See for example~: \textbf{Tang} \cite{tang_muon_2006}, \textbf{Shukla} \citep{shukla2018energy}, \textbf{Honda} \citep{Honda_2006qj}, \textbf{Gaisser} \citep{gaisser1990cosmic} ...
    \item \textit{Cosmic rays shower generators} which simulate extended air showers (EAS) from the primary cosmic rays down to the particles at ground. The most popular are : \textbf{PARMA} \citep{sato2015analytical}, \textbf{CRY} \citep{hagmann2007monte}, \textbf{MUPAGE} \citep{carminati2008atmospheric}, \textbf{MCEq}  \citep{fedynitch_mceq_2016}, \textbf{CORSIKA}  \citep{heck1998corsika} ...
\end{itemize}

The sensitivity of the technique relies on the model's accuracy, which should take into account the experimental conditions in the most realistic possible way. For instance, it has been shown in Jourde et al. \citep{jourde2016monitoring} and Tramontini et al. \citep{tramontini2019middle}, that atmospheric conditions (pressure and temperature) are strongly correlated with the muon flux. It is also clear that the geomagnetic field may play a significant role by deflecting charged particles towards the poles \citep{clay1928penetrating}, which leads to a decrease of the flux at the equator and an increase at high latitudes.\\

 This article details the methodology of a flux simulation based on CORSIKA, a very flexible generator which allows to change the atmosphere and geomagnetic field profiles in a given place and time. We configured the hadronic interaction models and necessary options to obtain an accurate flux fitting our experimental data. We also show how the the obtained model compares to various analytical models. The last sections present numerical results on the effects of geomagnetic field and atmospheric conditions (pressure, temperature ..) on the muon flux. A global conclusion is drawn on the obtained tool performance. 

\section{Simulation strategy and validation}

\subsection{CORSIKA's parameters}
CORSIKA (COsmic Ray SImulations for KAscade) is a Monte Carlo code for simulating atmospheric particle showers initiated by high energy cosmic ray particles. Primary particles (protons, light nuclei ...) are tracked in the atmosphere until they interact, decay or are absorbed. All secondary particles are explicitly followed along their trajectories. Their parameters are stored when they reach an observational level. For more details on the physics involved in atmospheric showers processes, see reference \cite{heck1998corsika}. In this study, we have selected CORSIKA's options that are believed to be essential to simulate the muon flux in a specific way and we detail some of them below.

\subsubsection{Hadronic interaction models and primary particles trajectories}
The primary CR flux  is composed of several types of particles (H, He, C, O, Fe ...). When a primary particle reaches the top of the atmosphere, it undergoes hadronic interactions leading to the production of secondary CR. Among those particles, pions and kaons decay into muons. Different types of primary particle interaction models are available on CORSIKA. For the hadronic interaction models, we chose FLUKA for the low-energy interactions and QGSJET-II-4 for high energies, the best candidates in their energy domains.\\

The energy range of the primary particle is chosen to match the muon energy measurable in our tomography experiments~: $10^E$ with E=[1, 7]. This total range is split into several parts with a defined step. The number of simulated primary particles is weighted in each bin with an empiric law for the number of showers~: $N=10^{9-E}$. This choice is also motivated by issues on computational time which increases considerably with the energy of the primary particle. The energy spectrum of primary CR follows a roughly exponential law : $E^{-\gamma}$. $\gamma$ is calculated for each intermediate energy range with a fit defined by an analytical model describing the primary spectrum. We select the Papini et al. model \citep{papini1996estimate} based on real data fits. This reference allows to take into account the solar modulation on the expected primary CR. The total flux is the sum of the fluxes of the various constituents of primary CR. It is possible to run simulations for each primary particle type or to apply the \textit{superposition model} explained by Spurio \citep{spurio2015cosmic} on the muon fluxes (results presented here). On top of these energy and momentum parameters, the trajectories of particles are defined by their zenith and azimuth angles, ranging from 0° to 90 ° and from -180° to 180°, respectively. For zenith angles higher than 60° the curvature of the atmosphere is taken into account, it cannot be neglected.

\subsubsection{Earth's magnetic field}
The Earth is protected by a magnetic shield created by the Earth's magnetosphere, which reduces the intensity of the high-energy flux reaching the ground. The geomagnetic field (B) modifies the spectrum of particles bombarding our atmosphere. This presents a low-energy cutoff, where the Earth's magnetic field is able to deflect primary CR below 10 GeV near the equator and close to 1 GeV at higher latitudes. The primary CR intensity also varies with longitude because of the asymmetry of the geomagnetic axis with respect to the Earth's rotation axis \citep{pethuraj2019measurement}. Those "East-West" fluxes show differences in energy intensity up to 100 GeV. This difference is more marked at high altitude than on the ground. Finally, there are significant local variations of the geomagnetic field, which affect the intensity of CR, the most famous being the South Atlantic Anomaly (SAA). All these effects can be taken into account in our simulation.\\

For each place, we declare the horizontal ($B_x$) and vertical ($B_z$) components of the Earth's magnetic field (in~$\mu$T), $B_x$ being the magnetic north. They are generated by \textit{NOAA geomagnetic calculator} according to the reference \citep{tapia_first_nodate}. CORSIKA computes the total magnetic field and its inclination from these two components. The main magnetic field effects are mostly latitude-dependent and therefore related to $B_x$, as we detail in section \ref{sectionB}. 

\subsubsection{Atmosphere properties}

The atmosphere can be divided into five layers: the troposphere, stratosphere, mesosphere, thermosphere and ionosphere. The atmospheric profiles presented in section \ref{section4} stop at the end of the stratosphere ( $\sim$ 50 km). The troposphere is the part of the Earth's atmosphere located between the surface and an altitude of about 8 to 15 kilometers, depending on latitude and season. It is thicker at the equator than at the poles. This layer concentrates three quarters of the atmospheric mass and the temperature decreases rapidly with altitude. The stratosphere extends, on average, between 12 and 50 km. It is characterized by an increase in temperature with altitude. The stratosphere begins at low altitude near the poles, because the temperature is lower there. The distribution of atmospheric density is therefore different at opposite latitudes.\\

Muons are produced at typical 10-15 km altitude (troposphere/stratosphere boundary). Their abundance is affected by the density differences in the atmosphere either by direct  re-interaction or by modification of their parent mesons survival probabilities before decay \citep{gaisser2016cosmic, grashorn2010atmospheric}. The effect is more important for high-energy muons, which result from high-energy mesons with larger lifetime due to time dilation and therefore with longer paths in the atmosphere. Thus high-energy muons are more sensitive to temperature changes. An input to CORSIKA is therefore the atmosphere's state in which the CR ray showers are generated. The state of the atmosphere is described by the density of the air at each altitude level. This one is calculated by converting the relative humidity into saturation vapor pressure with the Magnus formula \citep{abreu2012description}. We computed the parameters and altitudes of the layer boundaries from ERA5 data, the latest climate reanalysis produced by the ECMWF which combines large amounts of meteorological observations with estimates made from advanced modeling and data assimilation systems. Some atmospheric density profiles built are represented on Figure \ref{subplot_tout_T_density}.

\subsection{Differential fluxes from analytical models Vs CORSIKA}
\begin{figure}[h]
    \centering
  \includegraphics[scale=0.32,trim=1.4cm 0cm 1.5cm 0cm,clip=true]{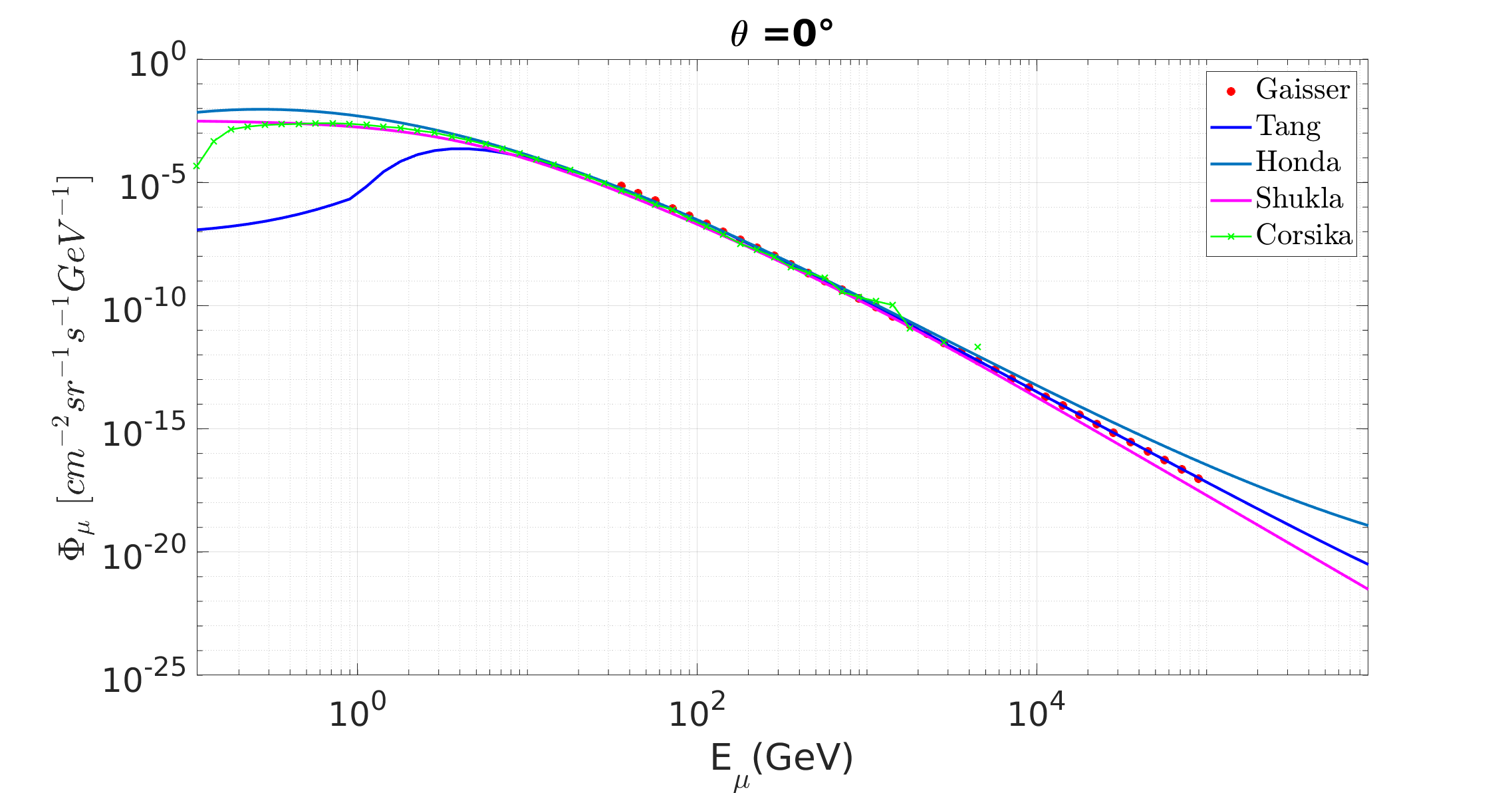}
    \caption{Differential fluxes as function of muon energy for a given zenith angle $\theta$=0°. Comparison between CORSIKA simulation and analytical models. }
    \label{corsikadiff}
\end{figure}

The fluxes simulated with CORSIKA, along the procedures described before, are compared to analytical fluxes to check their relevance. For this purpose, we plot differential fluxes as a function of muon energy only, for a muon zenith angle equal to 0°, on Figure \ref{corsikadiff}. At low energies they disagree : the models fit on several orders of magnitude to end up separated at high energies. But we know that low-energy muons are important for calibration and high-energy muons for conducting tomography experiments. Analytical models are known to be poorly adapted to small and large energies, because few measurements are available for their fitting equations. The CORSIKA model instead is probably more reliable over the whole energy range. Furthermore, analytical models are not extrapolated for all zenith angles and they don’t take into account geodesics parameters, a limitation overcome by the CORSIKA approach. 

\subsection{Comparison of a CORSIKA flux with real data  \label{corsikavsdata}}
\begin{figure}[h]
    \centering
  \includegraphics[scale=0.26,trim=1.4cm 0cm 1.5cm 0cm,clip=true]{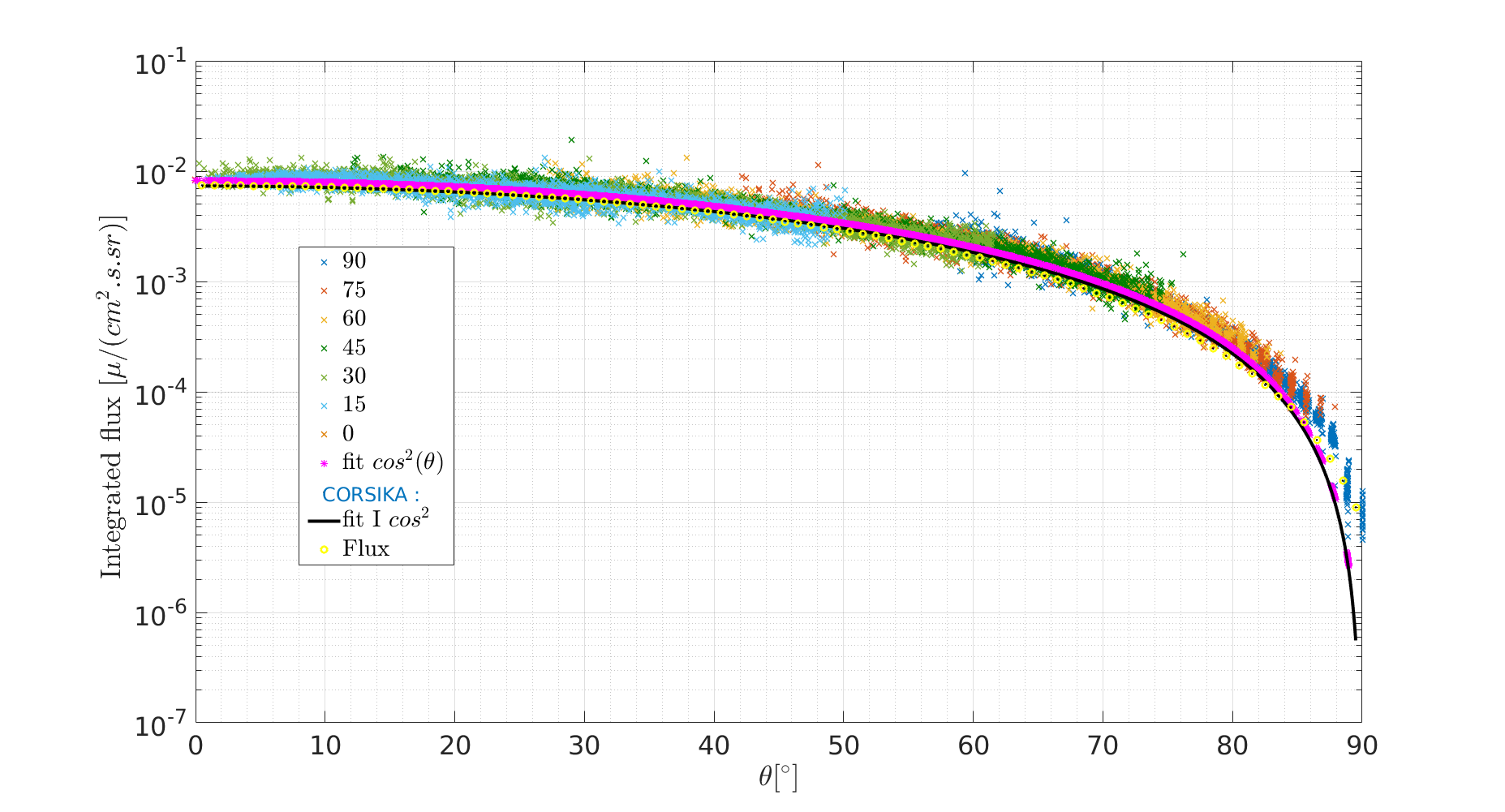}
    \caption{Integrated muon fluxes plotted as a function of the zenith angle $\theta$. The colored crosses represent the measured data for different inclinations of the detector. The pink crosses are a fit in $cos^2(\theta)$ on the total measured flux. The circles represent the simulated flux with the CORSIKA. A $cos^2(\theta)$ fit on it is drawn with a solid black line.}
    \label{crossvalidation}
\end{figure}

The best validation cross-checks for any simulation is the comparison to real data. Data presented here were taken in Lyon (France, latitude ~45° and close to sea level), in open sky conditions, with a 3 planes muon tracker (so-called muon telescope). We tilted the telescope progressively by step of $\ang{15}$ from the zenith ($\theta=\ang{0}$) to the horizontal ($\theta=\ang{90}$) directions. The muon flux is simulated in Lyon (France) respecting the geodesic constraints. Figure \ref{crossvalidation} displays the data/simulation comparison. Experimental points are the small crosses (1 tilt = 1 color) and data from CORSIKA's simulations are represented by yellow circles. Different fits are made on these two datasets with a simple $cos^2(\theta)$ at first order. Despite the still pending disagreement at large zenith angles, we observe a real improvement in the data/model comparison with respect to the usual $cos^2(\theta)$ (or analytical) fit. The simulated flux under-evaluates slightly the measured one for high zenith angles. To overcome this, it is necessary to constrain the simulated flux to the data to be as close as possible to reality. 

\section{Numerical Results: effects of the geomagnetic field}\label{sectionB} 
In this section, we used our CORSIKA model, validated on a large dataset to simulate muon fluxes (E, $\theta$) for various 
Earth magnetic field. We wanted to quantify those effects on the measurable muon flux in open sky. For this purpose, the atmospheric and altitude parameters have been fixed to constant values. 

\subsection{$B_x$ influence} 
\begin{figure}[h]
    \centering
\includegraphics[scale=0.33,trim=1.5cm 0cm 1.5cm 0cm,clip=true]{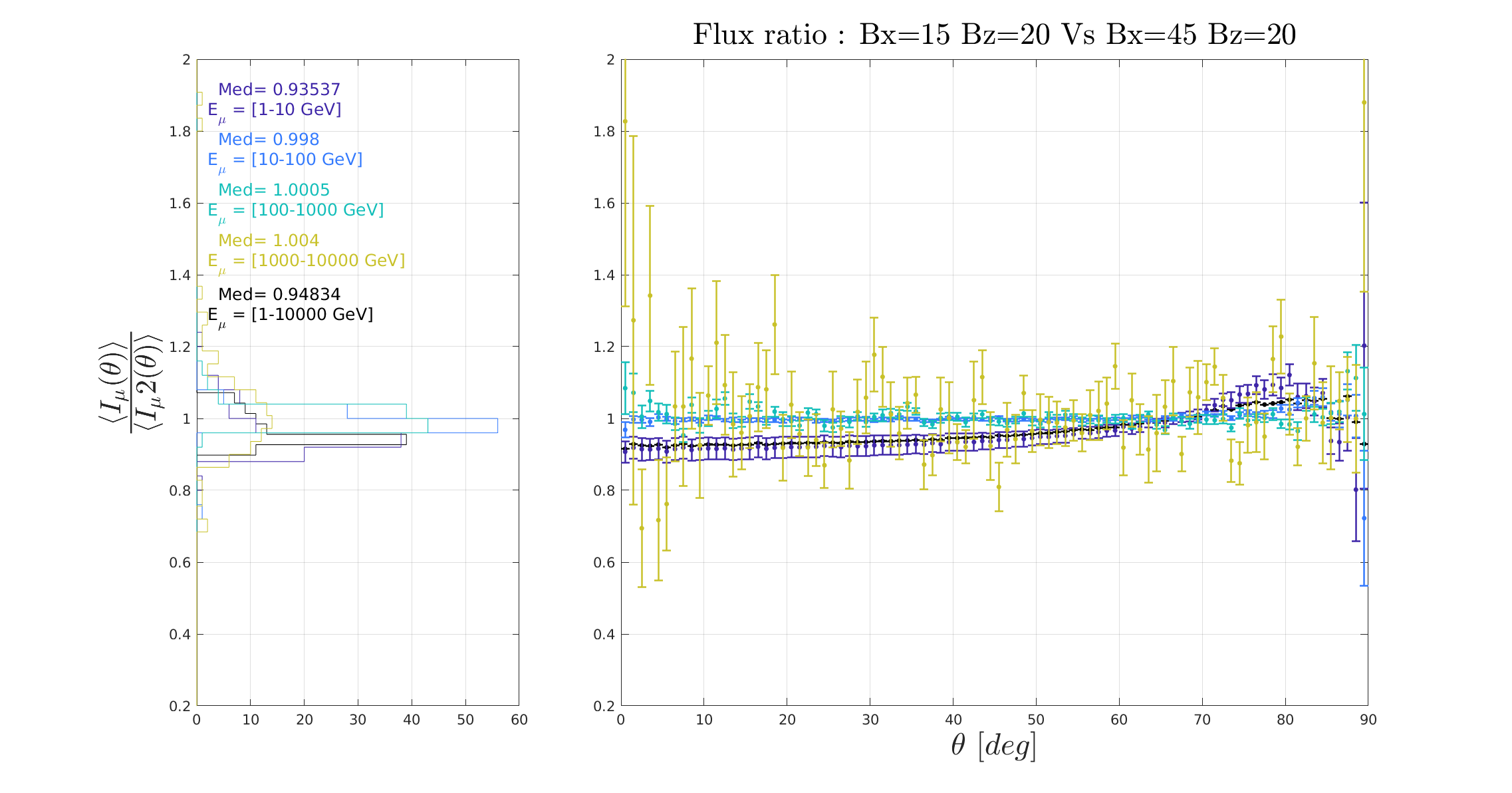}
\includegraphics[scale=0.33,trim=1.5cm 0cm 1.5cm 0cm,clip=true]{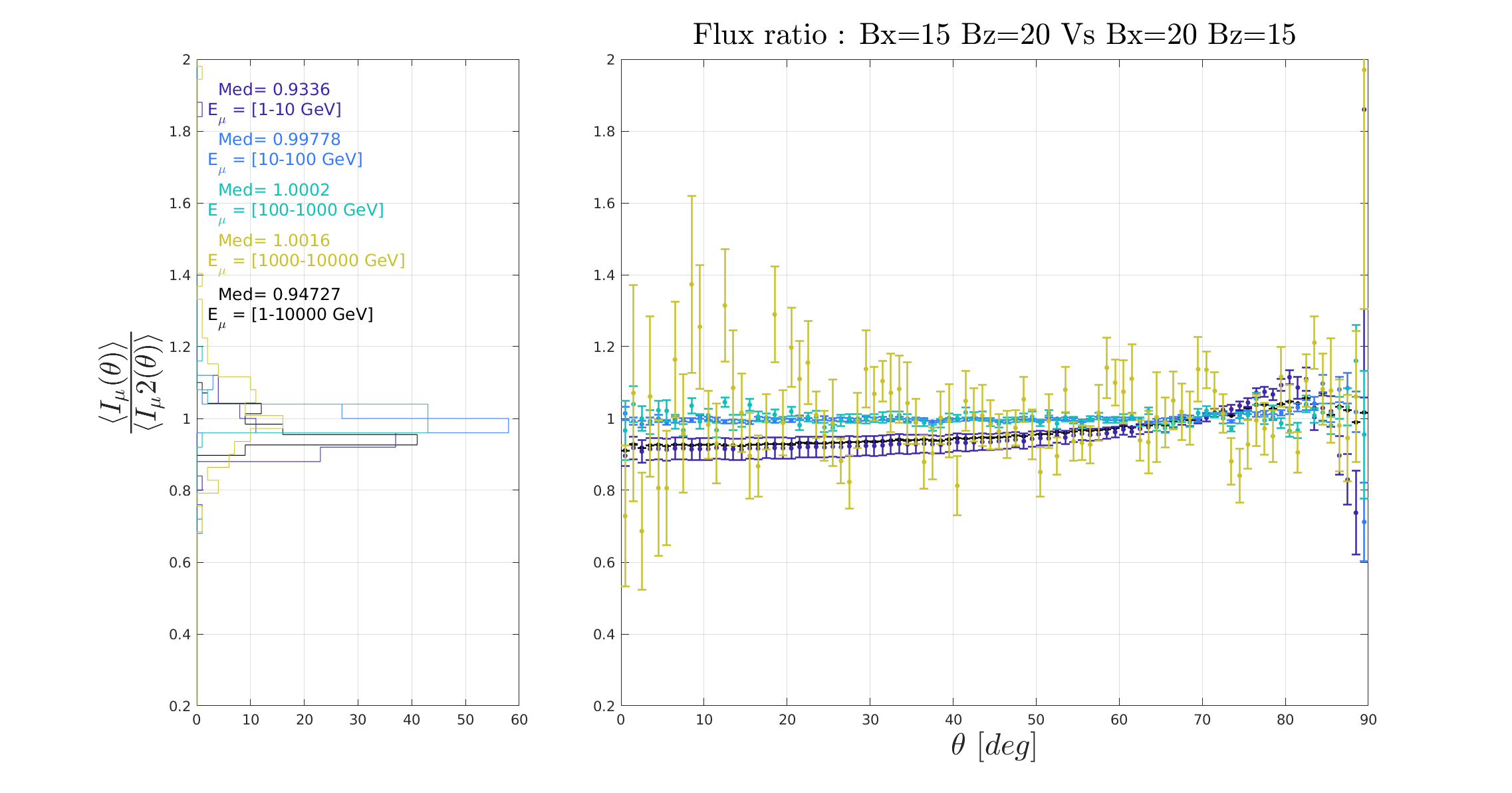}
    \caption{Comparison of two simulations, with same atmosphere and altitude, and different magnetic fields. (Top) The magnetic parameters are (1) $B_x$ = 15 $\mu$T, $B_z$ = 20 $\mu$T and (2) $B_x$ = 45  $\mu$T, $B_z$ = 20 $\mu$T. (Bottom) The magnetic parameters are : (1) $B_x$ = 15  $\mu$T  $B_z$ = 20  $\mu$T and (2) $B_x$ = 20  $\mu$T  $B_z$ = 15  $\mu$T. 
    (Left panel) Normalized intensity distribution for different energy ranges from 1 to 10$^4$ GeV (in color) and for the whole energy range (in black). (Right panel) Flux ratio dependence on the zenith angle $\theta$ between two different magnetic field fluxes for the same energy ranges as in the left panel.}
    \label{$B_x$15BZ20$B_x$45BZ20_1GeV}
\end{figure}

We have set the vertical component as a constant ($B_z$= 20 $\mu$T) and the horizontal component ($B_x$ = 15 $\mu$T) for a simulation of reference. We then compute another simulation with parameters set at $B_x$ = 45 $\mu$T, $B_z$ = 20 $\mu$T. We simulate the muon flux with those two configurations and compute their ratio for four different energy ranges and different zenith angles. Figure \ref{$B_x$15BZ20$B_x$45BZ20_1GeV} (top), on left panel, presents the normalized intensity distribution of the flux ratio over the zenith angle, for each individual energy range. 
It shows that when considering a larger energy range, the intensity ratios distribution tends to a narrow peak centered at 0.94834 corresponding to an effect of about 5$\%$. As expected the ratio tends to one for energy ranges greater than 10 GeV but affects the low-energy particles which are more deflected towards the poles. Right panel shows the flux ratio for different zenith angles $\theta$ for the same energy ranges. At low energies (1 to 10 GeV in purple) the flux is higher for $B_x$= 45 $\mu$T until 60° and between 85° and 90°, and lower between 60 and 85°. This probably arises from the fact that high-angles particles cross a larger section of the atmosphere and may be less deflected. For higher energies, the flux ratio remains more constant at each angle (in light blue, green and yellow). The total integrated flux ratio (in black) follows the behaviour of the "1 to 10 GeV" integrated flux.\\

\subsection{$B_z$ influence}

The same procedure is followed this time with a fixed horizontal component ($B_x$ = 30  $\mu$T) and various horizontal components ($B_z$~=~20 and 45  $\mu$T). We simulate the muon flux with these two configurations and we compute the ratio as a function of the energy and of the zenith angle. The results are not shown in this issue because only a slight difference at high zenith angles is observed. The $B_z$ component does not seem to affect the muon flux. \\

\subsection{$B$ intensity influence}

Here we have kept the previous simulation of reference, and compute another simulation with where the $B_x$ and $B_z$ components values have been switched. The intensity for the total geomagnetic field has been kept constant for both simulations (Figure \ref{$B_x$15BZ20$B_x$45BZ20_1GeV} (bottom)).
In this case the conclusions are the same as for Figure  \ref{$B_x$15BZ20$B_x$45BZ20_1GeV}. Low-energy muons are more deflected when $B_x$ increases. It is precisely this component that influences the muon flux because in this subsection the magnetic field intensity remained constant in both fluxes compared. \\

These numerical results validate that the magnetic field particularly affects low-energy muons which are deflected towards the poles. The very energetic mesons are less affected by the geomagnetic field. Low-zenith angle muons are less affected because the trajectory is short, at high zenith angle it is less clear because they are less and less energetic as they progress.\\

All the results presented in this section are subject to significant uncertainty. It is statistical and increases when the energy increases. Indeed, we simulate much less high-energy muons and extreme-angles muons. We have limits imposed by computational time and linked to the randomness of CORSIKA. 

\section{Numerical results : Atmospheric parameters influence} \label{section4}

We desire to quantify the impact of atmospheric density on the flux, which is ultimately controlled by the temperature and pressure state of the atmosphere. Hence, at the same location, seasonal variations affect the muon flux over time. 
For our tests we chose the city of Lyon in France (lat = 45.75, long = 4.75, $B_x$ = 22.71  $\mu$T, $B_z$ = 40.96  $\mu$T), place where we performed the open sky measurements shown in Figure \ref{crossvalidation}
The atmosphere density parameters used to simulate muon fluxes with CORSIKA are determined with "era5tool" and ERA5 datasets, for two different dates.

\subsection{Atmosphere models}
\begin{figure}[H]
    \centering
  \includegraphics[scale=0.33,trim=1.5cm 0cm 1.5cm 0cm,clip=true]{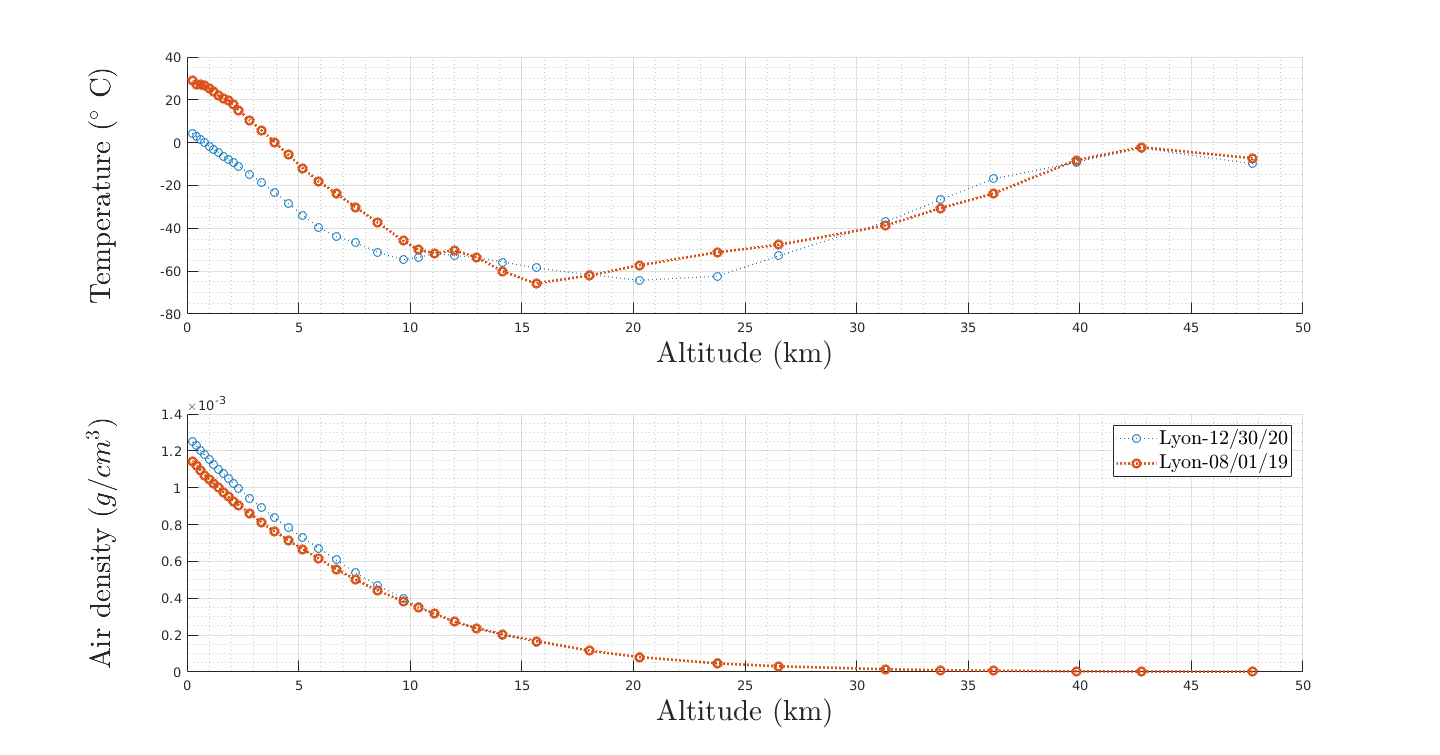}
    \caption{(Top) Temperature and (Bottom) density profiles for altitudes from 0 to 50 km in Lyon during winter (30/12/20) in blue and in summer (08/01/19) in red.}
    \label{subplot_tout_T_density}
\end{figure}

Temperature and density profiles in Lyon, during winter (12/30/20) and summer (08/01/19), are displayed on Figure \ref{subplot_tout_T_density}. These temperature and density profiles highlight which part of the atmosphere may affect the muon flux production and filtering. We observe that the density of the atmosphere decreases with altitude, and that colder atmospheres are denser especially at lower altitude. 

\subsection{Atmosphere flux comparison : Seasonal effects}

\begin{figure}[H]
    \centering
  \includegraphics[scale=0.33,trim=1.5cm 0cm 1.5cm 0cm,clip=true]{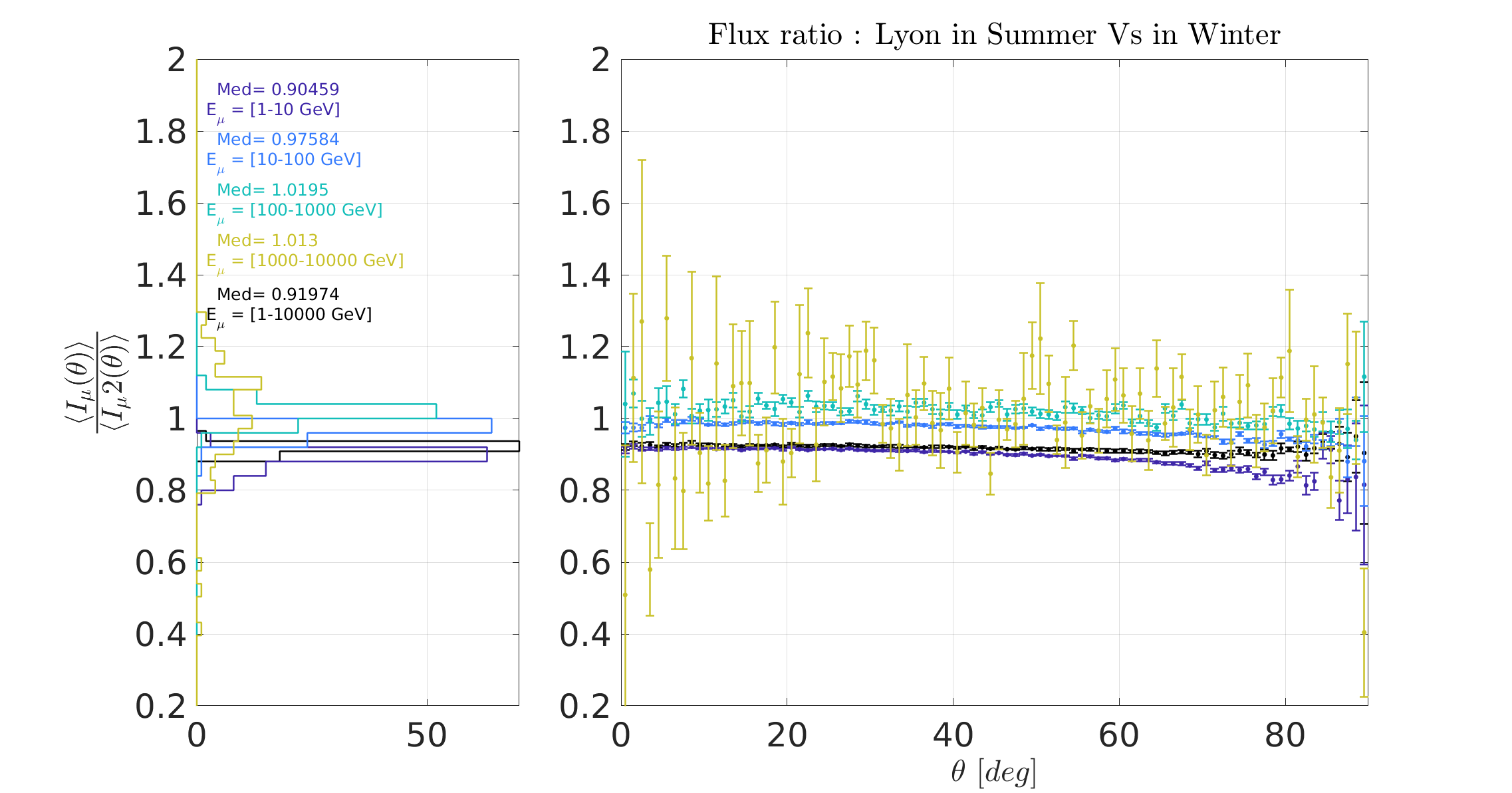}
    \caption{(Left) Normalized intensity distribution for several energies range from 1 to 10 000 GeV (in color) and for the whole energy range (in black). (Right) Flux ratio as a function of the zenith angle theta in summer and winter conditions in Lyon, France for the same energy ranges as left panel. The different atmosphere conditions are (1) 08/01/2019, (2) 12/30/20, with constant geomagnetic field and altitude.}
    \label{lyon010819-301220_intflux}
\end{figure}

Figure \ref{lyon010819-301220_intflux} presents the normalized intensity distribution for different energy ranges : 1-10 GeV, 10-100 GeV,  100-10$^3$ GeV, 10$^3$-10$^4$ GeV and the distribution for the whole energy range. It shows that when considering a larger energy range, the intensity distribution tends to a narrow peak centered at 0.91974 which means that the flux is 10 $\%$ higher in winter in Lyon (France). This effect is quite sizeable and must be properly accounted for, when precise muography is required over for low opacity targets. This effect depends of course on the particles energy. Right panel shows the flux ratio for different zenith angles $\theta$ in the range 0 to 90°, when considering the same energy ranges. It shows that in summer the flux is higher for high energies (100 to 10$^4$ GeV, in green and yellow), and lower in winter. For lower energies, the flux ratio is higher in winter (1 to 100 GeV, in light and dark blue), and lower in summer. These effects increase with the zenith angles $\theta$. Statistical errors of fluxes simulations are present on Figure \ref{lyon010819-301220_intflux} as in Section \ref{sectionB}. \\

 The decrease in atmospheric density increases the muon flux. We have seen before that the effect is more important for high-energy muons, which result from high-energy mesons with larger lifetime due to time dilation and therefore with longer paths in the atmosphere. We observe this effect on Figure \ref{lyon010819-301220_intflux} where the flux increases by 1$\%$ during summer. 

\section{Conclusion}

In this paper, we present a muon flux simulation workflow accounting for muon-atmosphere interactions, based on the CORSIKA framework. We detailed our simulation strategy and the various relevant inputs from the hadronic interactions models to the atmosphere conditions. In particular, we used meteorological ERA5 pressure and temperature datasets to compute the required atmospheric density profiles. The workflow has been cross-validated against experimental evidences and standard semi-empiric models found in the literature. Simulations prove themselves to be a powerful tools to study and make predictions on tiny effects induced by the geomagnetic field or the atmospheric seasonal variations. Those effects are of increasing importance when one wants to produce muon imagery, and/or, long-term internal state surveys, on both ends of the opacity spectrum. Low-opacity targets imagery is controlled by low-energy muons filtered out by the density of the atmosphere. On the other side, high opacity targets imagery is largely affected by the process at stake for high-energy muon production. This study opens the gate to develop semi-empiric formulas predicting the evolution of the muon energy spectrum for each zenith angle, in relation with the atmospheric state. These formulas will be useful to correct recorded muon fluxes on the fly, when direct open-sky measurements are not available or not sufficiently refined in terms of energy description.  

\section*{Acknowledgements}
This work was the subject of a CIFRE agreement between the ArcelorMittal Maizières Research SA and IP2I (Lyon).\\

\vspace{2cm}
\color{black}
\noindent\rule[0.25\baselineskip]{\textwidth}{1pt}
\bibliography{biblio}

\end{document}